\documentclass[10pt]{article}
\usepackage{wrapfig}
\usepackage{color}
\usepackage{graphicx,framed}
\usepackage[colorlinks=true, pdfstartview=FitV, linkcolor=blue, citecolor=blue, urlcolor=blue]{hyperref}
\definecolor{shadecolor}{rgb}{0.95, 0.95, 0.86}

\usepackage{amsbsy}
\usepackage{amssymb}

\def\Id{\mathrm {Id}}
\def\Ai{ {\rm Ai}}
\usepackage{verbatim}
\makeatletter
\@addtoreset{equation}{section}
\makeatother

\newtheorem{theorem}{Theorem}[section]

\newtheorem{exercise}{Exercise}[section]

\newtheorem{lemma}{Lemma}[section]
\newtheorem{remark}{Remark}[section]

\newtheorem{proposition}{Proposition}[section] 
\newtheorem{corollary}{Corollary}[section] 
\newtheorem{definition}{Definition}[section]
\def\le{\left}
\def\ri{\right}

\def\br{\begin{remark}}
\def\er{\end{remark}}
\def\bt{\begin{theorem}}
\def\et{\end{theorem}}
\def\bc{\begin{corollary}}
\def\ec{\end{corollary}}
\def\bx{\begin{examp}\small}
\def\ex{\end{examp}}
\def\bxr{\begin{exercise}\small}
\def\exr{\end{exercise}}
\def\bl{\begin{lemma}}
\def\el{\end{lemma}}
\def\bd{\begin{definition}}
\def\ed{\end{definition}}
\def\bp{\begin{proposition}}
\def\ep{\end{proposition}}
\def\be{\begin{equation}}
\def\ee{\end{equation}}
\def\&{\hspace{-15pt}&}
\def\bea{\begin{eqnarray}}
\def\eea{\end{eqnarray}}
\def\beas{\begin{eqnarray*}}
\def\eeas{\end{eqnarray*}}

\def\C{{\mathbb C}}

\def\R{{\mathbb R}}

\def\wh{\widehat}

\def\d{\,\mathrm d}

\def\l{\lambda}

\def\1{{\bf 1}}

\date{}
\begin{document}
\baselineskip 15pt plus 1pt minus 1pt

\vspace{0.2cm}
\begin{center}
\begin{Large}
\fontfamily{cmss}
\fontsize{17pt}{27pt}
\selectfont
\textbf{On the location of poles for the Ablowitz-Segur  family of solutions to the second Painlev\'e\ equation}
\end{Large}\\
\bigskip
\begin{large} {M.
Bertola}$^{\ddagger,\sharp}$\footnote{Work supported in part by the Natural
    Sciences and Engineering Research Council of Canada (NSERC).}\footnote{bertola@mathstat.concordia.ca}
\end{large}
\\
\bigskip
\begin{small}
$^{\ddagger}$ {\em Department of Mathematics and
Statistics, Concordia University\\ 1455 de Maisonneuve W., Montr\'eal, Qu\'ebec,
Canada H3G 1M8} \\
$^{\sharp}$ {\em Centre de recherches math\'ematiques\\ Universit\'e\ de Montr\'eal } \\
\end{small}
\bigskip
{\bf Abstract}
\end{center}
Using a simple operator-norm estimate we show that the solution to the second Painlev\'e\ equation within the Ablowitz-Segur family is pole-free in a well defined region of the complex plane of the independent variable. The result is illustrated with several numerical examples.

\vspace{0.7cm}

\section{Introduction and result}

Since the introduction of the Painlev\'e\ equations and in recent years, due to their appearance in many different models of mathematical physics there has  been a growing interest in the ``Painlev\'e\ program'' of classifying solutions, connection formul\ae\ etc. (see the book \cite{ItsKapaevFokasBook} as a prime example of this program). 

The six Painlev\'e\ equations (and all the other equations with the Painlev\'e\ property) are such that the only  ``movable'' singularities of the general solution\footnote{By ``movable'' it is meant a singularity whose position depends on the member of the family, e.g. on the initial values.} are poles, whereas all the ``nastier singularities'' (essential singularities, branchpoints) are fixed (they depend on  the particular equation under scrutiny only). For example for the first two equations 
\be
(PI) \ \ u''(s) = 6u(s)^2 -s\ ,\ \ \ \ (PII) \ \ u''(s) = su(s) +2 u(s)^3\ ,
\ee
each solution is meromorphic, with $s=\infty$ being the essential singularity. 
Given a special solution to a Painlev\'e\ equation, there is a certain interest in determining the location of its poles; within this circle of ideas we may mention the work of \cite{KapaevP1} where it is shown that the {\em tritronqu\'ee} solution to Painlev\'e\ I has no poles in the sector $|\arg (s)|<4\pi/5$ and $|s|$ sufficiently large. Following this, and due to its relevance in the semiclassical asymptotics of the Nonlinear Schr\"odinger equation, Dubrovin conjectured \cite{DubrovinGravaKlein} that --in fact-- the {\em tritronqu\'ee} is pole-free in the whole sector (not just the distal part).
Asymptotic methods \cite{Masoero} and numerics based on Pad\'e\ approximants \cite{Novokshenov-Pade} provide solid evidence that the aforementioned conjecture holds true. 

The study of the  distribution of poles for special solutions is in general an elusive task; in very special cases, by using the correspondence with an appropriate Riemann--Hilbert problem, one may use certain ``vanishing lemmas'' if the corresponding problem has sufficient symmetry (typically a Schwartz-reflection symmetry of some sort, see for example \cite{Ablowitz-Fokas-book, FokasZhouP2}). These techniques allow at best to prove that certain solutions are pole-free on the real axis, for example. 
When venturing into the complex plane of the independent variable, the required symmetry is lost and those methods cannot be applied.

Our more modest goal here is to study the pole distribution of the so-called  Ablowitz--Segur family \cite{Ablowitz-Segur} of solution to Painlev\'e\ II;
\bea
u''(s)  &\& = s u(s) + 2 u^3(s),\label{P2ODE}\\
&\& u(s) \simeq \kappa Ai(s) \ ,\qquad s\to + \infty, \ \kappa\in \C.\label{asympt}
\eea

A particular 
and very important
 example of this family corresponds to the Hastings-McLeod solution $\kappa=1$; this solution has the property \cite{Hastings-McLeod} that it is pole-free for {\em real values} of the variable $s$; this property is maintained for $|\kappa|<1$ while for $\kappa\in \R \setminus [-1,1] $ the solution has poles on the real axis (but is still pole-free for sufficiently large positive $s$).

The Ablowitz-Segur family corresponds to purely imaginary solutions with behaviour as in (\ref{asympt}) but $\kappa \in i\R$; such solutions are oscillatory as $s\to - \infty$ and decay as $s\to +\infty$ and are also pole-free on the real axis.

In both the Hastings-McLeod and Ablowitz-Segur cases (and even more general solutions) a great deal of results are known, including connection formul\ae\ and  asymptotic form of the solutions. In particular the following  is known (which we rephrase here in a form adapted to the context of the present paper)\footnote{
The results in \cite{ItsKapaevFokasBook} are stated for even more general solutions of the general Painlev\'e II  ODE $u'' = s u + 2 u^3 + \alpha$ (our case being $\alpha=0$). Much more detail is contained in the cited  comprehensive book: some of those results appeared first in separate papers of the same authors and the specific references are contained ibidem.
 }
\bt[See Thm. 11.1, 11.7 in \cite{ItsKapaevFokasBook}]
\label{thm111}
For each $\kappa\in \C$ there exists a constant $R>0$ such that  the solutions of (\ref{P2ODE}) with behavior \ref{asympt} are pole-free in the region sector $\arg s\in \le[-\frac \pi 3, \frac \pi 3\ri]\ , \ \ |s|>R$. In the special case $\kappa=1$ (Hastings-McLeod solution) the region is $\arg(s)\in \le[-\frac \pi 3,\frac \pi 3 \ri]\cup \le[\frac {2\pi}3, \frac {4\pi}3\ri]\ ,\ \ \ |s|>R$. 
\et

The question arises as to how large is the region in the {\em complex} $s$--plane where the solution remains free of poles; this question is  -in principle- not easy to settle as the equation (\ref{P2ODE}) has the {\bf Painlev\'e\ property}, that is, its solutions have only poles as singularities, but their position depend on the initial data ({\em movable singularities}).
The result we shall prove is stated here:
\begin{shaded}
\bt
\label{main}
The solutions to $u''(s) = s u(s) + 2u^3(s)$ with the behavior 
\be
u(s) \simeq \kappa\, Ai(s) \ ,\qquad s\to + \infty, \ \kappa \in \C
\ee
are pole free in the {\em whole} region (the fractional power is the principal one)
\be
\Re (s^\frac 32) >\frac 32  \ln |\kappa|\label{region}
\ee
In particular if $|\kappa|=1$ the region coincides with the sector $\arg s \in (-\pi/3,\pi /3)$.
\et
\end{shaded}
\br
In general the negative real axis is excluded since $s^\frac 32 \in i\R$ for $s\in -\R$ (no matter which determination of the fractional power); however, if $|\kappa|<1$ or if $\kappa \in i\R$ it is known that there are no poles on the negative axis. See later Remark \ref{improved}. We also can replace the inequality (\ref{region}) by   $\Re (s^\frac 32)\geq \frac 32  \ln |\kappa|$, see Remark \ref{improved2}.
\er
\section{Proof of Thm. \ref{main}}
The proof is based upon the following
\bp
\label{propdet}
The solution of (\ref{P2ODE}) with the asymptotics (\ref{asympt}) is the second logarithmic derivative of the Fredholm determinant of the integral operator on $H:=L^2(\gamma_+ \cup \gamma_-, |\d \l|)$ with kernel
\be
K_s(\lambda,\mu):= \frac {{\rm e}^{\frac i 2 (\theta_s(\lambda) - \theta_s(\mu))} \le[ \chi_{+}(\lambda) \chi_-(\mu) +  \chi_{+}(\lambda) \chi_-(\mu)\ri]  } {2i\pi(\lambda - \mu)}\ ,\ \ \ \theta_s(\l):= \frac {\l^3}3 + s \l,
\ee
where $\chi_{\pm}(\lambda)$ denote the indicator functions of the sets $\gamma_\pm := \R\pm i c$, $c>0$ (with $\gamma_+$ oriented from right to left and $\gamma_-=-\gamma_+$ oriented from left to right), as follows 
\be
u(s)^2  = -\frac {\d ^2}{\d s^2} \ln \det \bigg(\Id - \kappa K_s\bigg)
\ee
\ep
The proof is contained in \cite{BertolaCafasso1}; it essentially is a different representation of the Airy-operator\footnote{Relative to loc. cit. we have performed an overall rotation $\l\mapsto i\l$ so that the phase function that was $\vartheta(\l):= \frac {\l^3}3 - x \l$ becomes the one used in the present paper $\theta_x(\l) := \frac {\l^3}3 + x \l$ (with an overall factor of $-i$). Therefore the contours that were denoted $\gamma_L, \gamma_R$ (for ``left'' and ``right'') have been rotated to $\gamma_+$ and $\gamma_-$, respectively.} on $L^2(\R,\d x)$ with integral kernel
\be
K_{{\rm Ai}}(x,y) := \frac {\Ai(x)\Ai'(y)- \Ai'(x)\Ai(y)}{x-y} = \frac {i}{4\pi^2} \int_{\gamma_-}\!\!\!\! \d \mu \int_{\gamma_+} \!\!\!\! \d\l \frac { {\rm e}^{ i\theta_y(\l) - i\theta_x(\mu)}}{\mu-\l}
\ee
Indeed it is shown in [\cite {BertolaCafasso1}, Thm. 3.1] that (in the present notation)
\be
\det\le(\Id_{L^2(\gamma_+\cup \gamma_-)} - \kappa K_s\ri) = \det  \le(\Id_{L^2( [s\,\infty))} - \kappa^2 K_{{\rm Ai}}\bigg|_{[s,\infty)}\ri)
\label{deteq}
\ee
where the right-hand side is   the Fredholm determinant that appears in the spacing distributions of the Gaussian Unitary Ensemble (GUE) of random matrices \cite{TracyWidomLevel}
\be
F_2(s;\kappa):= \det \le(\Id_{L^2( [s\,\infty))} - \kappa^2 K_{{\rm Ai}}\bigg|_{[s,\infty)}\ri)\label{TW}
\ee
and it is  related to the Ablowitz-Segur family (\ref{P2ODE}, \ref{asympt}) by 
\be
u(s;\kappa)^2:= -\frac {\d^2}{\d s^2} \ln F_2 (s;\kappa).
\ee
The simple idea behind  the proof of identity (\ref{deteq}) is that of re-expressing its restriction to the semi-interval $[s,\infty)$ in Fourier space. 
The advantage is that, while on one hand  it is unclear how to extend the representation (\ref{TW}) to complex values of $s$, on the other hand the kernel $K_s$ of Prop. \ref{propdet} depends analytically on  $s$. 
\br
The determinant of the above operator is independent of the details of the contours $\gamma_\pm$; the usual choice for $\gamma_+$ would be any (smooth)  contour in the upper half plane extending to infinity in such a  way that $\|{\rm e}^{i\theta_s}\| = \mathcal O(|\l|^{-\infty})$. Typically this contour extends along directions $\arg \l = \frac \pi 6, \frac {5\pi}6$, but we can use $\R+ic$, $c>0$ thanks to the simple computation 
\be
\Re\,\le[  i (x+ic)^3\ri]= -3c x^2  + c^3.
\ee 
\er
\bc
The poles of the solutions $u(s)$ mentioned in Prop. \ref {propdet} coincide with the zeroes of the Fredholm determinant  $\det \bigg(\Id - \kappa K_s\bigg)$ as a function of $s$.
\ec
It follows at once that {\em if the operator norm of $\kappa K_s$ can be estimated above by unity} in certain regions of the $s$--plane, the Fredholm determinant is therefore nonvanishing and the corresponding solution is pole-free; and this is precisely what we set out to accomplish below.

We shall follow \cite{BertolaCafasso1} and  represent the operator $K_s$\footnote{We abuse the notation and denote the operator by the same symbol as its integral kernel.} as a block-antidiagonal operator relative to the natural splitting $L^2(\gamma_+\cup \gamma_-) = L^2(\gamma_+)\oplus L^2(\gamma_-)$, and then  use the following identity of Fredholm determinants
\be
\det\le[\Id_{L^2(\gamma_+\cup \gamma_-)} - \kappa \le[
\begin{array}{c|c}
0 & \mathcal G\\
\hline
\mathcal F & 0
\end{array}
\ri] \ri] = \det \le [\Id_{L^2(\gamma_-)} - \kappa^2 \mathcal G\circ \mathcal F\ri]
\label{DetId}
\ee
where the two operators $\mathcal F, \mathcal G$ are defined as follows; 
\bea
L^2(\gamma_-) \mathop{\longrightarrow \atop \longleftarrow}^{\mathcal F}_{\mathcal G}L^2(\gamma_+)\\
(\mathcal F g)(\lambda):=  {\rm e}^{- \frac i 2\theta_s(\lambda)}  \int_{\R-ic} \frac {\d\mu}{2i\pi}
 \frac { {\rm e}^{\frac i 2\theta_s(\mu)} g(\mu)} {\mu-\l} \\
(\mathcal G h)(\mu):={\rm e}^{\frac i 2\theta_s(\mu)}  \int_{\R+ic} \frac {\d\lambda}{2i\pi} \frac { {\rm e}^{- \frac i 2\theta_s(\lambda)} h(\lambda)}{\mu-\l} 
\eea
The identity (\ref{DetId}) holds because (as proven in \cite{BertolaCafasso1}) both $\mathcal F, \mathcal G$ are of trace-class in $L^2(\gamma_+)\oplus L^2(\gamma_-)$.
It is also clear that $\mathcal F$ and $\mathcal G$ (between the respective spaces) have the same norms by the symmetry; we shall now make a very rough (but sufficient) estimate of their norms, from which the (simple) proof of Thm. \ref{main} shall follow at once.

We shall consider $\mathcal F$ for definiteness, and we start by observing that  it is the composition of three operators:
\begin{itemize}
\item the multiplication by ${\rm e}^{\frac i 2 \theta_s(\mu)}$ on $L^2(\R-ic)$;
\item the Cauchy operator from $L^2(\R+ic)$ to $L^2(\R-ic)$;
\item the  multiplication by ${\rm e}^{-\frac i 2 \theta_s(\mu)}$ on $L^2(\R +  ic)$.
\end{itemize}
The norm of the Cauchy operator 
\be
\mathcal C:L^2(\R - ic)\mapsto L^2(\R+ ic)
\ee
is promptly shown  to be unity. To see this fact,  let $\mu=x+ic$ and $\l = y-ic$, so that the above operator can be written out simply as an operator of $L^2(\R, \d x)$,
\be
(\mathcal C f)(y) := \frac 1{2i\pi} \int_\R \frac {f(x)\d x}{x-y  -  2ic}
\ee
Since this is a convolution operator with the function $\varphi(x) = \frac {1}{2i\pi(-2ic -x)}$, its representation in Fourier transform\footnote{We use the convention $\hat f(t) = \frac 1{\sqrt{2\pi}} \int_\R {\rm e}^{-it x} f(x) \d x$.} is the multiplication operator by $\sqrt{2\pi} \wh \varphi(t)$ where  the Fourier transform of the function  $\varphi(x)$ is easily computed with the aid of the residue theorem to be 
\be
\hat \varphi(t) = \frac 1{\sqrt{2\pi}} {\rm e}^{-2tc} \chi_{\R_+}(t).
\ee
Since the norm of a multiplication operator is its $L^\infty$ norm and the Fourier transform is unitary, we conclude that $\||\mathcal C\|| = 1$.

Thus the norm of $\mathcal F$ (and similarly $\mathcal G$) is bounded by the $L^\infty$ norm of the multiplication operator; to achieve the optimal estimate we must choose appropriately $c$ so that the line $\R+ic$ passes through the saddle point of the phase function $\theta_s = \frac  {\l^3}3 +s \l$. Elementary calculations give that the saddle point is $ \l_s:=\pm  i \sqrt{s}$, $\arg (s)\in (-\pi,\pi]$. We thus need to choose $c_0:=\Re \sqrt s = \sqrt{|s|}\cos\le(\frac {\arg(s)}2\ri)$. We leave to the reader the elementary verification that 
\bea
\max_{\l\in \R - ic_0} \le| {\rm e}^{\frac i2 \theta_s(\l)}\ri| = {\rm e}^{-\frac 1 3 \Re (s^{\frac 32})}\ ,\qquad
\max_{\mu \in \R + ic_0} \le| {\rm e}^{-\frac i2 \theta_s(\mu)}\ri| = {\rm e}^{-\frac 1 3 \Re (s^{\frac 32})}\ ,
\eea
and that the maxima are taken on at the saddle points only.
It follows thus that 
\be
|\| \mathcal F\|| \leq  \|  {\rm e}^{\frac i2 \theta_s(\l)} \|_{L^\infty(\R- i c_0)} 
 \|  {\rm e}^{\frac i2 \theta_s(\mu)} \|_{L^\infty(\R+ i c_0)}  =  {\rm e}^{-\frac 2 3 \Re (s^{\frac 32})}
\ee
A completely parallel computation shows that 
\be
|\| \mathcal G\|| \leq   {\rm e}^{-\frac 2 3 \Re (s^{\frac 32})}
\ee
and hence 
\be
|\|\kappa^2 \mathcal G\circ \mathcal F\|\ \leq |\kappa|^2  {\rm e}^{-\frac 4 3 \Re (s^{\frac 32})} \label{mainineq}
\ee
Thus, if the right side of the inequality (\ref{mainineq}) is bounded by $1$ the operator $\Id - \kappa^2  \mathcal G\circ \mathcal F$ is invertible and thus the Fredholm determinant is nonzero.  The condition $ |\kappa|^2  {\rm e}^{-\frac 4 3 \Re (s^{\frac 32})} <1$ is precisely the condition of Thm. \ref{main}, which is thus proved.

\br
\label{improved}
The operator $\mathcal G \circ \mathcal F$ is basically the Fourier transform of the kernel $\chi_{[s,\infty)}(x) K_\Ai(x,y)$; as such it is shown in \cite{TracyWidomLevel} that its norm (for $s\in \R$, of course!) is strictly less than one and hence the Fredholm determinant (\ref{DetId}) cannot vanish if $|\kappa|\leq 1$. Moreover it is well known  that $K_{\Ai}$ on $L^2(\R)$ is a positive operator and hence for $\kappa\in i\R$ we also have that the determinant of $\Id + |\kappa|^2 K_\Ai$ cannot vanish.
\er

\begin{figure}
\begin{tabular}{cc}
\begin{tabular}{c}
\resizebox{0.3\textwidth}{!}{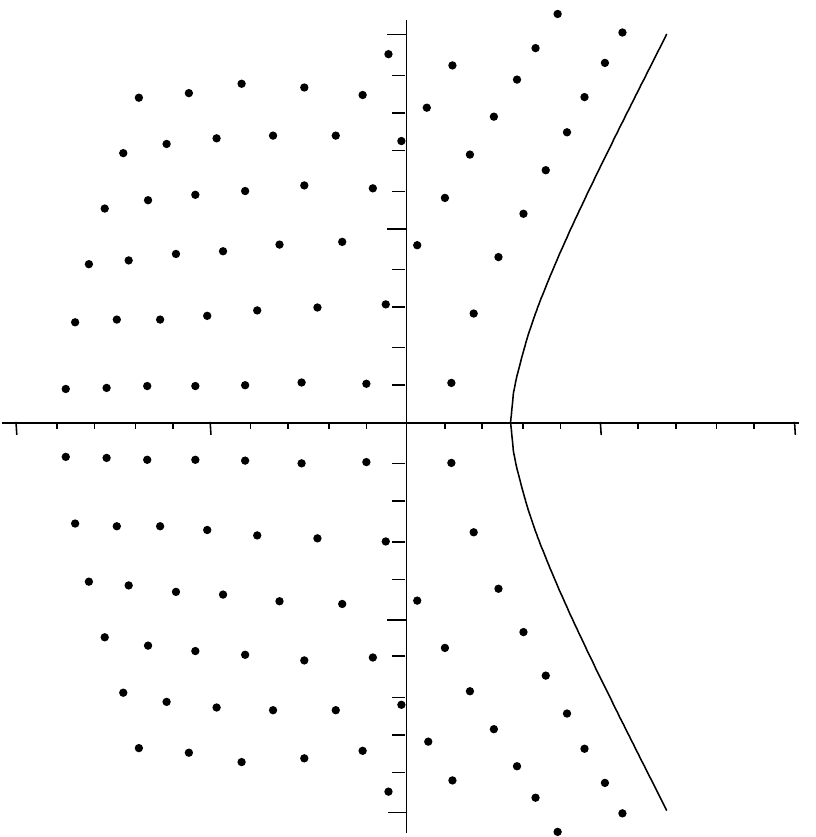}\\
{$\kappa = 20\,i$}
\end{tabular} &
\begin{tabular}{c}
\resizebox{0.4\textwidth}{!}{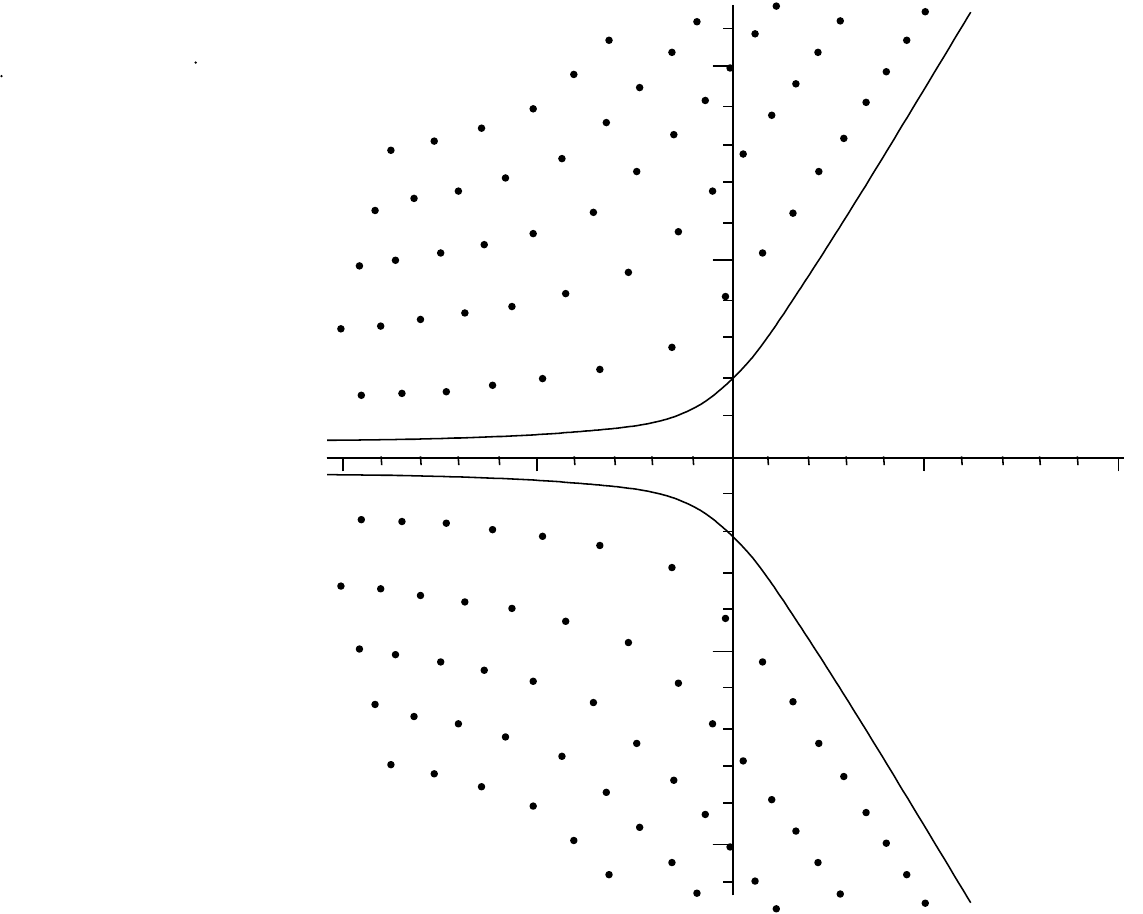}\\
$\kappa = 0.3\, i$
\end{tabular}\\
\begin{tabular}{c}
\resizebox{0.35\textwidth}{!}{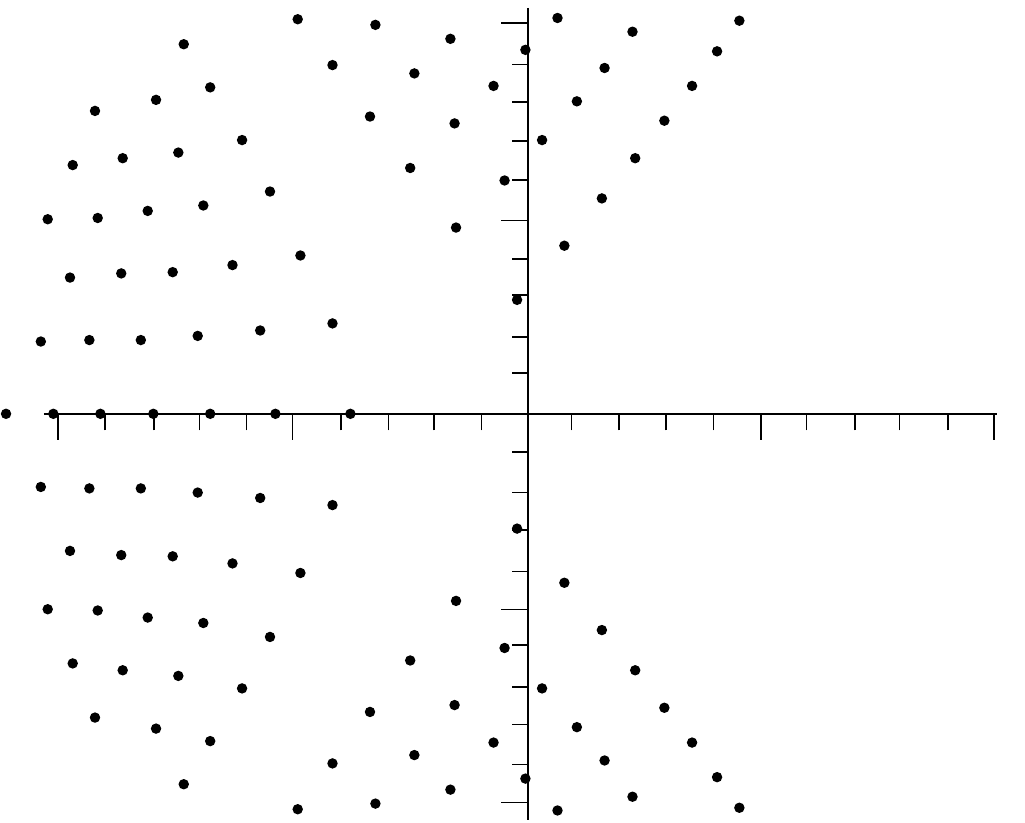}\\
$\kappa = 1.01$
\end{tabular}
& 
\begin{tabular}{c}
\resizebox{0.35\textwidth}{!}{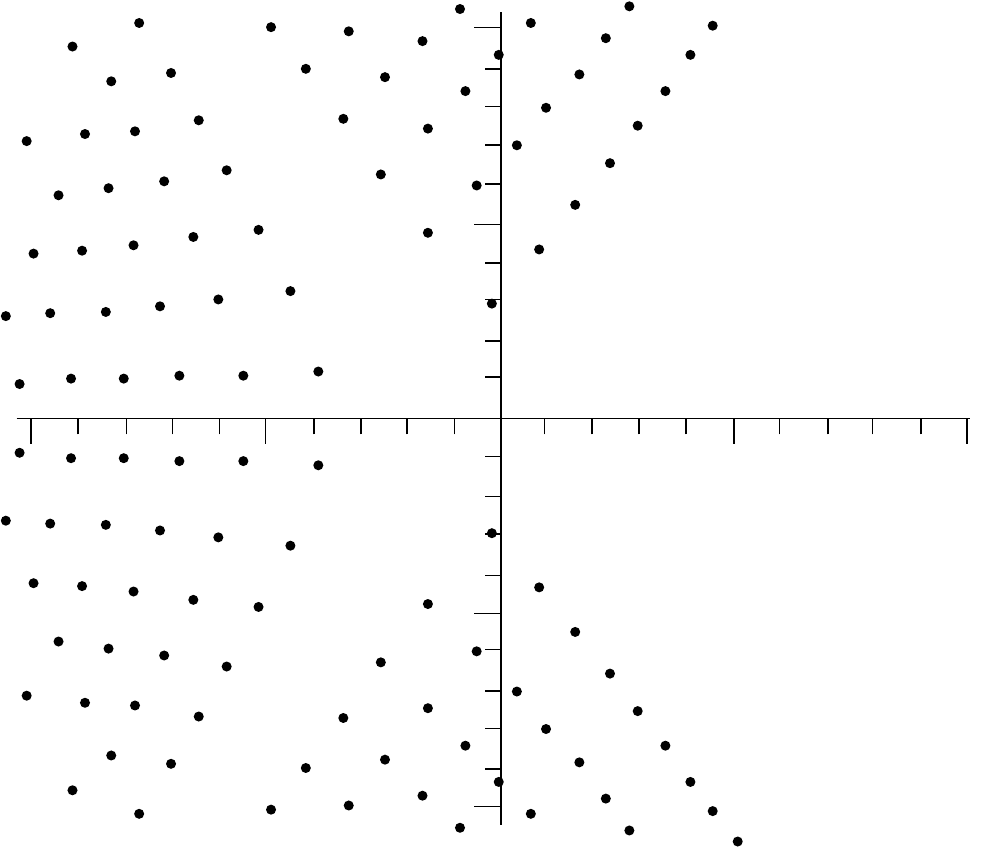}\\
$\kappa = 0.99$
\end{tabular}\\
\begin{tabular}{c}
\resizebox{0.35\textwidth}{!}{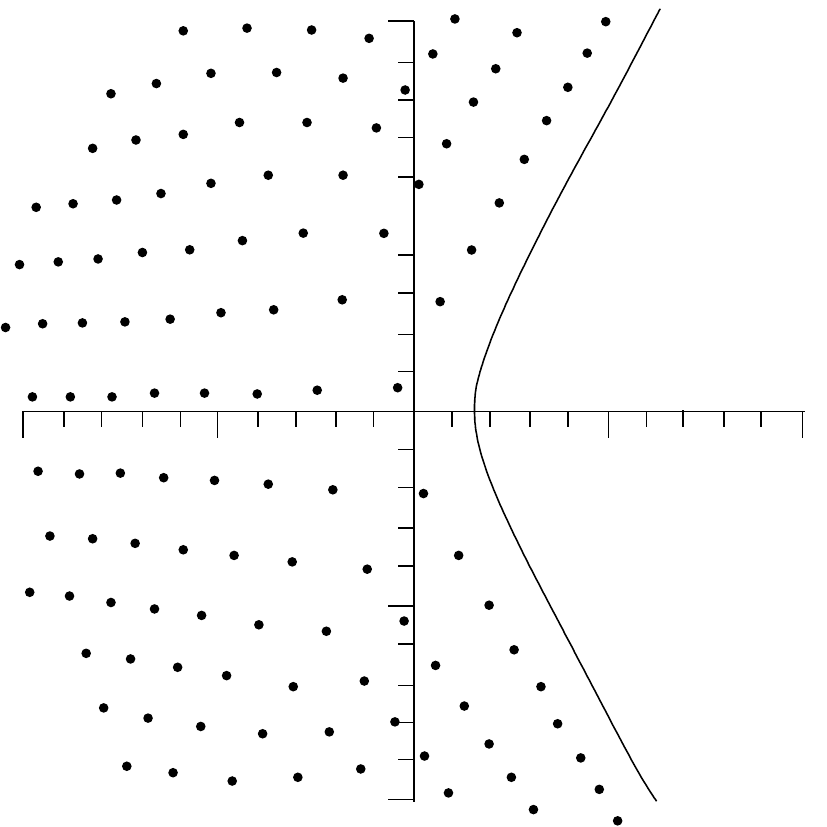}\\
$\kappa=3 + 2i$
\end{tabular}
&
\begin{tabular}{c}
\resizebox{0.35\textwidth}{!}{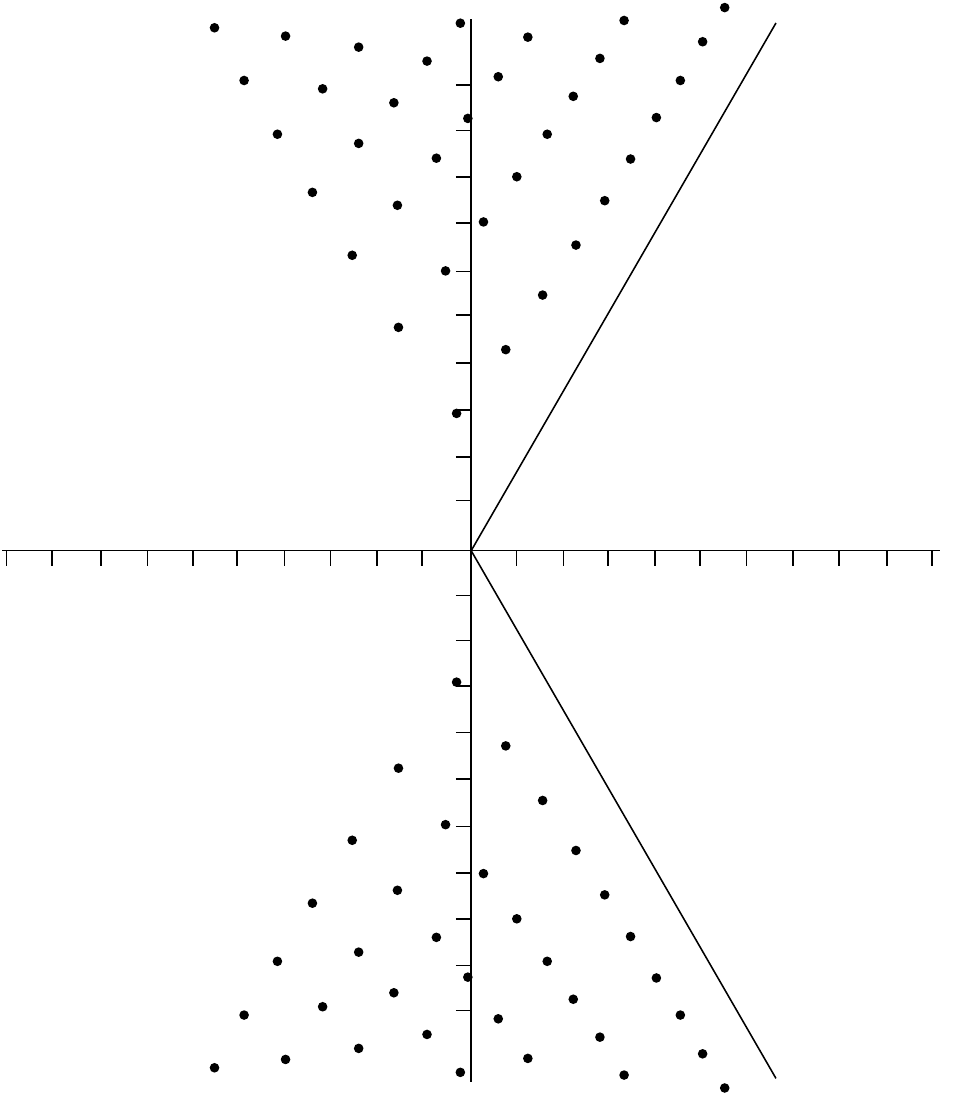}\\
$\kappa=1$ (Hastings-McLeod)
\end{tabular}
\end{tabular}
\caption{The poles of the Ablowitz-Segur solution with various indicated values of $\kappa$ and the boundary of the pole-free region as per Thm. \ref{main}. The numerics have been produced using the algorithm explained in \cite{Novokshenov-Pade} based on Pad\'e\ approximants to the solution. It also shows that the poles on the left tend to minus infinity as $\kappa$ tends to the critical value $\kappa=1$ (of course this is only a visual cue, not any proof). }
\label{numerics}
\end{figure}

\br
\label{improved2} 
Since the operator $\mathcal G\circ \mathcal F$ is trace-class, it has only discrete spectrum. Tracing the inequalities about the norms we see that when $c>0$ the Cauchy operator has only the null-space and then has only continuous spectrum, because in Fourier space it is a multiplication operator by the  function ${\rm e}^{-2tc}$, $t>0$ (nowhere constant). It then follows that the spectrum of $\mathcal G\circ \mathcal F$ is {\bf strictly} bounded  by the value in  (\ref{mainineq}) (the maximum eigenvalue cannot achieve the bound since it is at the boundary of the essential spectra of the various operators involved).
Therefore we can improve the statement of Theorem \ref{main} by replacing the strict inequality by  $\geq$.
\er

\br
The kernel of Prop. \ref{propdet} is an ``integrable kernel'' in the sense of \cite{ItsIzerginKorepinSlavnov} and it is thus immediately related to Riemann--Hilbert problem for a $2\times 2$ matrix; this problem turns out to be the very standard RHP in the isomonodromic approach to Painlev\'e\ II, see for example \cite{Its:2002p9}. Thus the estimate of the kernel $K_s$ is nothing but an estimate of the norm underlying singular-integral operator associated to said RHP. We could have phrased this paper in that language but we opted for the current presentation in the interest of brevity because it does not require introducing the setup of any Riemann--Hilbert problem.
\er

\section{Comments and conclusion}
Although very simple, the result of Thm. \ref{main} seems to be not fully appreciated in the literature, to the knowledge of the author. It is also clear from the numeric evidence in Fig. \ref{numerics} that the result is almost optimal; one may improve it only by a more careful estimate of the operators involved, which seems to me a rather difficult task. 

We would also like to briefly comment on the special member $\kappa=1$, the so--called Hastings-McLeod solution \cite{Hastings-McLeod}; again numerical evidence strongly suggests  that there are no poles also in the {\em whole} sector $\arg(s) \in [2\pi /3, 4\pi/ 3]$ 
 as remarked in the introduction (see the results of \cite{ItsKapaevFokasBook} summarized in Thm. \ref{thm111}), this is known for sufficiently large $|s|$ but only conjectured (and observed numerically) for $|s|$ finite \cite{Novokshenov-Pade}. We can offer an informal argument below that indicates that one is not to expect an estimate bounding the spectrum of $K_s$ within an a-priori domain that excludes the point $1$.

The inspection of the numerics (for example at the top right frame in Fig. \ref{numerics} ($\kappa = 0.3i$)), implies that the eigenvalues of the operator $K_s(\l,\mu)$ are not confined within the unit disk as $s$ ranges in the (left) sector $\arg s \in \le[\frac {2\pi}3, \frac {4\pi}3\ri]$. To see this recall that the poles that are clearly visible in the left sector of the indicated piture  correspond to at  least one eigenvalue (possibly with multiplicity) taking the value $\frac 1 \kappa$, which is larger than one in modulus. 

On the other hand for real $s\in \R_-$ the operator $K_s$ has the same spectrum of the Airy kernel because it is unitarily equivalent to it:  the results of \cite{TracyWidomLevel} 
imply that {\em all} the eigenvalues $\l_j(s)$ are (simple and) confined in the interval $[0,1)$.  Therefore, by continuity, there is a region around the negative axis where all the eigenvalues are confined within the unit disk; on the other hand the aforementioned numerical evidence indicates that these eigenvalues exit the unit disk as $s$ ranges in the left sector.  In view of the above,  one could expect that there are values of $s$ in the left sector  at which some of the eigenvalues equal one: the fact that this is not the case, for $|s|$ large is implied by Thm. \ref{thm111}, and for arbitrary $|s|$ is suggested by the numerics.
In conclusion, a proof of the absence of poles in the whole sector cannot follow from simple estimates directly on the norm of the operator $K_s$ alone 
  and is probably much more delicate to achieve.
%
%
\paragraph{Acknowledgements:}
The author would like to thank A. Tovbis for illuminating discussions.
\bibliographystyle{plain}
\def\cprime{$'$}

\end{document}